\documentclass[prl,aps,amssymb,twocolumn]{revtex4}
\usepackage[dvips]{epsfig,color}
\usepackage{graphicx}
\usepackage{amsmath,amsfonts,amsbsy,amssymb}

\begin{document}

\title{Supersymmetric Quantum Hall Effect on Fuzzy Supersphere}

\author{Kazuki Hasebe}
\affiliation{Yukawa Institute for Theoretical Physics, Kyoto University,
 Kyoto 606-8502, Japan \\
Email: hasebe@yukawa.kyoto-u.ac.jp}

\begin{abstract}
Supersymmetric quantum Hall liquids are constructed on a supersphere
in a supermonopole background. 
We derive a supersymmetric generalization of the Laughlin wavefunction, which is a ground state of a hard-core $OSp(1|2)$ invariant Hamiltonian.
 We also present excited topological objects, which are fractionally charged deficits made by super Hall currents.
Several relations between quantum Hall systems and their  supersymmetric extensions are discussed.  
\end{abstract}

\maketitle

%%%%%%%%%%%%%%%%%%%%%%%%%%%%%%%%%%%%%%%%%%%%%%%%%%%%%%%%%%%%%%%%%%%%%%%%%%%%%%%%%%%%%%%%%%%%%%%%%%%%%%%%%%%%%%%%%%%%%%%%%%%%%%%%%%%%%%%%%%%%%%%%%%%%
%\section{Introduction}
%%%%%%%%%%%%%%%%%%%%%%%%%%%%%%%%%%%%%%%%%%%%%%%%%%%%%%%%%%%%%%%%%%%%%%%%%%%%%%%%%%%%%%%%%%%%%%%%%%%%%%%%%%%%%%%%%%%%%%%%%%%%%%%%%%%%%%%%%%%%%%%%%%%%

 Quantum Hall systems contain noncommutative structures, like Matrix theories and D-brane physics \cite{hep-th/0101029%,hep-th/0103013
}.
They are perhaps the simplest known physical set-up of  
noncommutative geometry and exhibit  many of its exotic properties \cite{GirvinPhysRevB29%, cond-mat/9407031, hep-th/0209198
}.
Therefore, quantum Hall systems, which can be readily investigated in the laboratory, represent a practical alternative to the physics which string theory attempts to describe, which are still far beyond our realm of experimental capability.
This is one reason that  quantum Hall systems are very  fascinating.  
It is expected that ideas developed by investigating   quantum Hall systems will help in furthering our understanding of the  high energy physics \cite{hep-th/0210162}.

Originally,  quantum Hall phenomena were realized in two-dimensional
 flat space under strong magnetic field.
Laughlin derived a wavefunction that well describes  
quantum  incompressible liquids \cite{LaughlinPRL50}.
 His wavefunction  is  rotationally symmetric  but 
not translationally symmetric on the plane.
 Hence, it does not possess all the symmetries 
 in a  plane and is not suited for computer simulations.
Haldane overcame this problem by  constructing  quantum Hall systems on two-spheres in a Dirac magnetic monopole background \cite{HaldanePRL51}.
He constructed a Laughlin-like wavefunction, which we call the Laughlin-Haldane wavefunction,  on the sphere that possesses   all the 
rotational symmetries of the two-sphere.   
The sphere used in  by  Haldane's analysis  is simply a fuzzy two-sphere.
Recently, Zhang and Hu  have succeeded in constructing four-dimensional quantum Hall systems in a  $SU(2)$ Yang monopole background 
\cite{cond-mat/0110572}.
The systems they consider are quantum Hall liquids on  fuzzy four-spheres and, intriguingly, possess  brane-like excitations. 
%Moreover, ``gravity'' arise as their massless edge excitations. 
Because Matrix theories can be used to describe higher-dimensional
 spaces and possess extended objects, their quantum Hall systems are the first discovered ``physical'' systems that exhibit behavior 
 similar to that described by   Matrix models.
Their theory has attracted much attention and has 
 been developed by many authors 
\cite{hep-th/0407007}.
In particular, on the basis of 
  fuzzy complex projective manifolds, Karabali 
and Nair 
 have generalized them into even higher-dimensional quantum Hall systems \cite{hep-th/0203264}.
Hasebe and Kimura, based on higher-dimensional fuzzy spheres, have found another way to generalize them for an arbitrary even number of  dimensions in 
colored monopole backgrounds \cite{hasebekimura0310274}.
In fact, such developments in the study of quantum Hall systems have  
 provided information that may be important in obtaining an understanding of 
D-brane physics. In particular, it has been reported that with 
 use of the Dirac-Born-Infeld action, the higher-dimensional fuzzy spheres in Matrix models can be  identified  with dielectric D-branes in colored monopole backgrounds \cite{hep-th/0402044}. 

Recently, it was found that  non-anticommutative (NAC) field theory is naturally realized on  D-branes in 
  R-R field or graviphoton backgrounds \cite{hep-th/0302109%,hep-th/0302078, hep-th/0305248
}.
Also, it has been shown that, in supermatrix models,  fuzzy superspheres arise as 
classical solutions, and  their fluctuations yield
  NAC field theories \cite{hep-th/0311005}.
Some interesting relations between  lowest Landau level (LLL) physics and NAC geometry have also been reported  \cite{hep-th/0306251%, hep-th/0311159
}.  
With these recent developments,
 it would be worthwhile  to extend the theory of 
quantum Hall systems to a  supersymmetric framework.
Indeed, the supersymmetric quantum Hall systems might be 
 the simplest ``physical'' set-up of NAC geometry.
 Further, encouraged by  previous success
 in the investigation of higher-dimensional quantum Hall systems,
we may hope  that such systems not only possess exotic properties 
in the NAC world but also  
reveal yet unknown aspects of supermatrix models.

\vspace{0.3cm}

%%%%%%%%%%%%%%%%%%%%%%%%%%%%%%%%%%%%%%%%%%%%%%%%%%%%%%%%%%%%%%%%%%%%%%%%%%%%%%%%%%%%%%%%%%%%%%%%%%%%%%%%%%%%%%%%%%%%%%%%%%%%%%%%%%%%%%%%%%%%%%%%%%%%
%\section{Spherical Super Quantum Hall Systems}
%%%%%%%%%%%%%%%%%%%%%%%%%%%%%%%%%%%%%%%%%%%%%%%%%%%%%%%%%%%%%%%%%%%%%%%%%%%%%%%%%%%%%%%%%%%%%%%%%%%%%%%%%%%%%%%%%%%%%%%%%%%%%%%%%%%%%%%%%%%%%%%%%%%
A supersphere is a geometrical object taking the form of a coset manifold   
   given by
$S^{2|2}=OSp(1|2)/U(1).$
By  construction, a supersphere manifestly possesses the exact  
 $\mathcal{N}=1$ supersymmetry, which is generated by the $OSp(1|2)$
 super Lie group.
The fact that the supersymmetry remains exact is an advantage of using the coset manifolds of  super Lie groups.
The number of degrees of freedom of the supersphere is given by
$dim S^{2|2}=dim OSp(1|2)-dim U(1)=5-1=4.$
Two of these degrees of freedom correspond to  the Grassmann even coordinates,    and 
the other two correspond to  the Grassmann odd coordinates on the supersphere.
The supersphere is embedded in a  flat superspace whose coordinates   are $x_a (a=1,2,3)$, which are Grassmann even, and $\theta_{\alpha} (\alpha=1,2)$, which are Grassmann odd. 
The radius of the supersphere is given by
$R=\sqrt{x_a^2+C_{\alpha\beta}\theta_{\alpha}\theta_{\beta}},$
where $C_{\alpha\beta}$ is an antisymmetric tensor with $C_{12}=1$.
At the center of the supersphere,  we place a supermonopole, whose magnetic charge  is  $I/2$ when $I$ is an integer. 
The supermonopole is useually referred as a graded monopole and is related to the supersymmetric extension of the 1-st Hopf map \cite{math-ph/9907020}.
The magnetic field of the supermonopole is given by 
$B={2\pi I}/{4\pi R^2}={I}/{2R^2}$,
and the magnetic length is defined as
$\ell_B\equiv {1}/{\sqrt{B}}={R}\sqrt{{2}/{I}}$.
The thermodynamic limit corresponds to $R, I \rightarrow \infty$,
  with $\ell_B$ fixed to  a finite value.
  For simplicity, in the following we set $R=1$, which makes all  quantities dimensionless.

%%%%%%%%%%%%%%%%%%%%%%%%%%%%%%%%%%%%%%%%%%%%%%%%%%%%%%%%%%%%%%%%%%%%%%%%%
%%%%%%%%%%%%%%%%%%%%%%%%%%%%%%%%%%%%%%%%%%%%%%%%%%%%%%%%%%%%%%%%%%%%%%%%%
%\section{One particle state}
%%%%%%%%%%%%%%%%%%%%%%%%%%%%%%%%%%%%%%%%%%%%%%%%%%%%%%%%%%%%%%%%%%%%%%%%%
%%%%%%%%%%%%%%%%%%%%%%%%%%%%%%%%%%%%%%%%%%%%%%%%%%%%%%%%%%%%%%%%%%%%%%%%%
We now briefly discuss the one-particle state on the  supersphere in
  the supermonopole background.
A detailed analysis is found in Ref.\cite{hasebekimura0409230}.
The one-particle Hamiltonian is given by
%%%%%%%%%%%%%%%%%%%%%%%%%%%%%%%%%%%%%%%%%%%%%%%%%%%%%%%%%%%%%%%%%%%%%%%%%
\begin{equation}
H=\frac{1}{2M}(\Lambda_a^2+C_{\alpha\beta}\Lambda_{\alpha}\Lambda_{\beta}),
\label{superQHhamiltonian}\nonumber
\end{equation}
%%%%%%%%%%%%%%%%%%%%%%%%%%%%%%%%%%%%%%%%%%%%%%%%%%%%%%%%%%%%%%%%%%%%%%%%%
where $\Lambda_a$ and $\Lambda_{\alpha}$ are the $OSp(1|2)$ covariant  particle  angular momenta, 
%%%%%%%%%%%%%%%%%%%%%%%%%%%%%%%%%%%%%%%%%%%%%%%%%%%%%%%%%%%%%%%%%%%%%%%%%
$\Lambda_a=-i\epsilon_{abc}x_bD_b+\frac{1}{2}\theta_{\alpha}(\sigma_a)_{\alpha\beta}D_{\beta}$ and $\Lambda_{\alpha}=\frac{1}{2}(C\sigma_a)_{\alpha\beta}x_aD_{\beta}-\frac{1}{2}\theta_{\beta}(\sigma_a)_{\beta\alpha}D_a$.
%%%%%%%%%%%%%%%%%%%%%%%%%%%%%%%%%%%%%%%%%%%%%%%%%%%%%%%%%%%%%%%%%%%%%%%%%
The covariant derivatives are given by  $D_a=\partial_a+iA_a$ and 
$D_{\alpha}=\partial_{\alpha}+iA_{\alpha}$, where  $A_a=-\frac{I}{2}\epsilon_{ab3}\frac{x_b}{1+x_3}
(1+\frac{2+x_3}{2(1+x_3)}\theta C\theta)$ and  
 $A_{\alpha}=-\frac{I}{2}i(\sigma_a C)_{\alpha\beta}x_a\theta_{\beta}$.
The supermonopole field strengths are given by 
$B_a=-\frac{I}{2}x_a$ and $B_{\alpha}=-\frac{I}{2}\theta_{\alpha}$.
%%%%%%%%%%%%%%%%%%%%%%%%%%%%%%%%%%%%%%%%%%%%%%%%%%%%%%%%%%%%%%%%%%%%%%%%%%
The commutation relations for the covariant angular momenta of the particle are obtained as 
%%%%%%%%%%%%%%%%%%%%%%%%%%%%%%%%%%%%%%%%%%%%%%%%%%%%%%%%%%%%%%%%%%%%%%%%%
$[\Lambda_a,\Lambda_b]=i\epsilon_{abc}(\Lambda_c-B_c)$,  
 $[\Lambda_a,\Lambda_{\alpha}]=\frac{1}{2}(\sigma_a)_{\beta\alpha}(\Lambda_{\beta}-B_{\beta})$ and 
$\{\Lambda_{\alpha},\Lambda_{\beta}\}=\frac{1}{2}(C\sigma_a)_{\alpha\beta}(\Lambda_a-B_a)$.
%%%%%%%%%%%%%%%%%%%%%%%%%%%%%%%%%%%%%%%%%%%%%%%%%%%%%%%%%%%%%%%%%%%%%%%%%%
Thus, they do not satisfy the $OSp(1|2)$ commutation relations exactly,  due to the presence of the supermagnetic field.
The total $OSp(1|2)$ angular momenta are  constructed as  
%%%%%%%%%%%%%%%%%%%%%%%%%%%%%%%%%%%%%%%%%%%%%%%%%%%%%%%%%%%%%%%%%%%%%%%%
$L_a=\Lambda_a+B_a$, and $L_{\alpha}=\Lambda_{\alpha}+B_{\alpha}$.
%%%%%%%%%%%%%%%%%%%%%%%%%%%%%%%%%%%%%%%%%%%%%%%%%%%%%%%%%%%%%%%%%%%%%%%%%
The operators $L_{\alpha}$ play the  role of the supercharge in the system.
The covariance under $OSp(1|2)$ transformations is expressed as 
%%%%%%%%%%%%%%%%%%%%%%%%%%%%%%%%%%%%%%%%%%%%%%%%%%%%%%%%%%%%%%%%%%%%%%%%%
$[L_a,X_b]=i\epsilon_{abc}X_c,[L_a,X_{\alpha}]=\frac{1}{2}(\sigma_a)_{\beta\alpha}X_{\beta}$ and $
\{L_{\alpha},X_{\beta}\}=\frac{1}{2}(C\sigma_a)_{\alpha\beta}X_a,$
%%%%%%%%%%%%%%%%%%%%%%%%%%%%%%%%%%%%%%%%%%%%%%%%%%%%%%%%%%%%%%%%%%%%%%%%%
where $X_a$ represents $L_a$,$\Lambda_a$ and $B_a$, and $X_{\alpha}$ represents $L_{\alpha}$,$\Lambda_{\alpha}$ and $B_{\alpha}$.

The $OSp(1|2)$ Casimir operator  for the total $OSp(1|2)$ angular momenta
$L_a$ and $L_{\alpha}$ are 
%%%%%%%%%%%%%%%%%%%%%%%%%%%%%%%%%%%%%%%%%%%%%%%%%%%%%%%%%%%%%%%%%%%%%%%%%
$L_a^2+C_{\alpha\beta}L_{\alpha}L_{\beta}=j(j+\frac{1}{2}),$
%%%%%%%%%%%%%%%%%%%%%%%%%%%%%%%%%%%%%%%%%%%%%%%%%%%%%%%%%%%%%%%%%%%%%%%%
where $j=\frac{I}{2}+n$. Here, $n=0,1,2,\cdots$ indicates the Landau level.
The supermonopole field is perpendicular to the surface of the supersphere, while the particle moves on the supersphere.
Therefore, the particle angular momenta are orthogonal to the supermonopole field: 
%%%%%%%%%%%%%%%%%%%%%%%%%%%%%%%%%%%%%%%%%%%%%%%%%%%%%%%%%%%%%%%%%%%%%%%%%%
$\Lambda_a B_a+C_{\alpha\beta}\Lambda_{\alpha}B_{\beta}= B_a\Lambda_a +C_{\alpha\beta} B_{\alpha} \Lambda_{\beta}=0.$
%%%%%%%%%%%%%%%%%%%%%%%%%%%%%%%%%%%%%%%%%%%%%%%%%%%%%%%%%%%%%%%%%%%%%%%%%%
Observing these relations, we find the energy eigenvalues  to be  
%%%%%%%%%%%%%%%%%%%%%%%%%%%%%%%%%%%%%%%%%%%%%%%%%%%%%%%%%%%%%%%%%%%%%%%%
$E_n=\frac{1}{2M}(n(n+\frac{1}{2})+I(n+\frac{1}{4}))$.
%%%%%%%%%%%%%%%%%%%%%%%%%%%%%%%%%%%%%%%%%%%%%%%%%%%%%%%%%%%%%%%%%%%%%%%%
Thus, in the LLL, the energy becomes 
%%%%%%%%%%%%%%%%%%%%%%%%%%%%%%%%%%%%%%%%%%%%%%%%%%%%%%%%%%%%%%%%%%%%%%%%
$E_{LLL}=\frac{1}{4}\omega,$
%%%%%%%%%%%%%%%%%%%%%%%%%%%%%%%%%%%%%%%%%%%%%%%%%%%%%%%%%%%%%%%%%%%%%%%%%
where $\omega=B/M$ is the cyclotron frequency.

The number of  unit cells, which each occupies an area $2\pi\ell_B^2$
 on the supersphere,  is 
%%%%%%%%%%%%%%%%%%%%%%%%%%%%%%%%%%%%%%%%%%%%%%%%%%%%%%%%%%%%%%%%%%%%%%%
$N_{\Phi}={4\pi}/{2\pi\ell_B^2}=I$.
%%%%%%%%%%%%%%%%%%%%%%%%%%%%%%%%%%%%%%%%%%%%%%%%%%%%%%%%%%%%%%%%%%%%%%%
It is convenient to define  the filling fraction as 
%%%%%%%%%%%%%%%%%%%%%%%%%%%%%%%%%%%%%%%%%%%%%%%%%%%%%%%%%%%%%%%%%%%%%%%%
$\nu=N/N_{\Phi}$.
%%%%%%%%%%%%%%%%%%%%%%%%%%%%%%%%%%%%%%%%%%%%%%%%%%%%%%%%%%%%%%%%%%%%%%%%
We now give a comment.
From the $OSp(1|2)$ representation theory, 
the dimension of the irreducible representation $j=I/2$ is
%%%%%%%%%%%%%%%%%%%%%%%%%%%%%%%%%%%%%%%%%%%%%%%%%%%%%%%%%%%%%%%%%%%%%%%%%
$D=(2j+1)+(2j)|_{j=I/2}=2I+1$.
%%%%%%%%%%%%%%%%%%%%%%%%%%%%%%%%%%%%%%%%%%%%%%%%%%%%%%%%%%%%%%%%%%%%%%%%%%
Therefore, the number of  states in the LLL is twice as large as  the number
 of  magnetic cells in the large $I$ limit.
This implies that in  each magnetic cell,
 there are two degenerate states due to the supersymmetry.
Hence, in this system, the value of the filling fraction $N/N_{\Phi}$ is twice as large 
as that in  the ordinary definition $N/D$.

%%%%%%%%%%%%%%%%%%%%%%%%%%%%%%%%%%%%%%%%%%%%%%%%%%%%%%%%%%%%%%%%%%%%%%%%%%%%%%%%%%%%%%%%%%%%%%%%%%%%%%%%%%%%%%%%%%%%%%%%%%%%%%%%%%%%%%%%%%%%%%%%%%%%
%\subsection{Super coherent states}
%%%%%%%%%%%%%%%%%%%%%%%%%%%%%%%%%%%%%%%%%%%%%%%%%%%%%%%%%%%%%%%%%%%%%%%%%%%%%%%%%%%%%%%%%%%%%%%%%%%%%%%%%%%%%%%%%%%%%%%%%%%%%%%%%%%%%%%%%%%%%%%%%%%%
The supercoherent state $\psi$ is defined as  
the state which is  aligned in the direction of   the supermagnetic flux, $(B_a,B_{\alpha}) \propto (x_a,\theta_{\alpha})$; that is, we have 
$l_a \psi \cdot x_a +C_{\alpha\beta}l_{\alpha}\psi\cdot \theta_{\beta}
=\frac{1}{2}\psi,$
where $l_a$ and $l_{\alpha}$ constitute the  fundamental representation of the $OSp(1|2)$ generators. Explicitly, these are 
$l_a=\frac{1}{2}
\begin{pmatrix}
&\sigma_a & 0\\
& 0   & 0 
\end{pmatrix}
\!\!,~
l_{\theta_1}=\frac{1}{2}
\begin{pmatrix}
& 0 & \tau_1 \\
& {\tau_2}^t & 0 
\end{pmatrix}
\!\!,~
l_{\theta_2}=\frac{1}{2}
\begin{pmatrix}
& 0  & {\tau}_2   \\
& -\tau_1^t  & 0   
\end{pmatrix}\!\!,$
where the quantities $\{\sigma_a\}$ are the Pauli matrices, while  $\tau_1=(1,0)^t$ and $\tau_2=(0,1)^t$.
Up to a $U(1)$ phase factor, the explicit form of the supercoherent state is found to be
$\psi=(u,v,\eta)^t
=(
\sqrt{\frac{1+x_3}{2}}(1-\frac{1}{4(1+x_3)}\theta C\theta),
\frac{x_1+ix_2}{\sqrt{2(1+x_3)}}(1+\frac{1}{4(1+x_3)}\theta C\theta ),
\frac{1}{\sqrt{2(1+x_3)}}((1+x_3)\theta_1+(x_1+ix_2)\theta_2))^t$.
%%%%%%%%%%%%%%%%%%%%%%%%%%%%%%%%%%%%%%%%%%%%%%%%%%%%%%%%%%%%%%%%%%%%%%%%%%
This is identical to  the super Hopf spinor,  which satisfies the relations 
$\psi^{\ddagger}l_a\psi=\frac{1}{2}x_a$ and $\psi^{\ddagger}l_{\alpha}\psi=\frac{1}{2}\theta_{\alpha}$,
where ${\ddagger}$ denotes the superadjoint, defined as $\psi^{\ddagger}=(u^{*},v^{*},-\eta^{*})$, and  $*$ denotes  pseudoconjugation, which acts on a  Grassmann odd number $\xi$ as $\xi^{**}=-\xi,~(\xi_1\xi_2)^{*}=\xi_1^{*}\xi_2^{*}$.
%(See  Ref.\cite{BookFrappat} for details.)
The coordinates $\{x_a,\theta_{\alpha}\}$ are  super-real, in the sense
 that we have  $(x_a^{*},\theta_{\alpha}^{*})=(x_a,C_{\alpha\beta}\theta_{\beta})$. 

The supercoherent state in a  supermonopole background can be  obtained
 similarly.
The supercoherent state, directed to 
the point $(\Omega_a,\Omega_{\alpha})$, 
 should satisfy the equation
$[\Omega_a(\chi)L_a+C_{\alpha\beta}\Omega_{\alpha}(\chi)L_{\beta}]\psi^{(I)}_{\chi}
(u,v,\eta)=-\frac{I}{2}\psi^{(I)}_{\chi}(u,v,\eta),$
where $\chi$ is a constant super Hopf spinor given by 
$\chi=(a,b,\xi)^t$,
which is mapped to the point 
$(\Omega_a,\Omega_{\alpha})$ on the supersphere  by 
$\Omega_a(\chi)=2\chi^{\ddagger}l_a\chi$ and $\Omega_{\alpha}(\chi)=2\chi^{\ddagger}l_{\alpha}\chi.$
%%%%%%%%%%%%%%%%%%%%%%%%%%%%%%%%%%%%%%%%%%%%%%%%%%%%%%%%%%%%%%%%%%%%%%%%%%
%%%%%%%%%%%%%%%%%%%%%%%%%%%%%%%%%%%%%%%%%%%%%%%%%%%%%%%%%%%%%%%%%%%%%%%%%
The supercoherent  state  is found to be,
$\psi^{(I)}_{\chi}(u,v,\eta)=(\chi^{\ddagger}\psi)^I=(a^{*}u+b^{*}v-\eta^{*}\xi)^I.$

%%%%%%%%%%%%%%%%%%%%%%%%%%%%%%%%%%%%%%%%%%%%%%%%%%%%%%%%%%%%%%%%%%%%%%%%%%%%%%%%%%%%%%%%%%%%%%%%%%%%%%%%%%%%%%%%%%%%%%%%%%%%%%%%%%%%%%%%%%%%%%%%%%%%
%\subsection{Super Cyclotron Orbits}\label{subseccyclo}
%%%%%%%%%%%%%%%%%%%%%%%%%%%%%%%%%%%%%%%%%%%%%%%%%%%%%%%%%%%%%%%%%%%%%%%%%%%%%%%%%%%%%%%%%%%%%%%%%%%%%%%%%%%%%%%%%%%%%%%%%%%%%%%%%%%%%%%%%%%%%%%%%%%%
Supermonopole  harmonics $u_{m_1,m_2}$ and $\eta_{n_1,n_2}$  are introduced on the supersphere. They
  form a  basis for  the LLL and  are eigenstates of  $L_z$ with the eigenvalues $\frac{m_2-m_1}{2}$ and $\frac{n_2-n_1}{2}$, respectively. 
Their explicit forms are, 
$ u_{m_1,m_2}=\sqrt{\frac{I!}{m_1 ! m_2 !}}u^{m_1}v^{m_2}, ~\eta_{n_1,n_2}=\sqrt{\frac{I!}{n_1 ! n_2 !}}u^{n_1} v^{n_1}\eta$,
where $m_1+m_2=I$ and $n_1+n_2=I-1$.
The degeneracy of $u_{m_1,m_2}$ is $(I+1)$, while that of   $\eta_{n_1.n_2}$ is $(I)$. 
Thus, the total degeneracy is $(2I+1)$, which is exactly the dimension of the Hilbert space of the LLL.
Thus, without including  any complex variables 
 $\{u^{*},v^{*},\eta^{*}\}$, the functions in  the LLL are constructed  from the variables $\{u,v,\eta\}$. 
For this reason, the $OSp(1|2)$ operators  are effectively represented as 
$L_a=\psi^t \tilde{l}_a \frac{\partial}{\partial\psi}$ and 
$L_{\alpha}=\psi^t \tilde{l}_{\alpha}\frac{\partial}{\partial\psi},$ 
where 
$\frac{\partial}{\partial \psi}=(
\frac{\partial}{\partial u},
\frac{\partial}{\partial v},
\frac{\partial}{\partial \eta})^t$ 
 and $\{\tilde{l}_a, \tilde{l}_{\alpha}\}$  forms a  complex representation of $OSp(1|2)$ with
$\tilde{l}_a= -{l_a}^*$ and ${\tilde{l}}_{\alpha}= C_{\alpha\beta}l_{\beta}$.  
%%%%%%%%%%%%%%%%%%%%%%%%%%%%%%%%%%%%%%%%%%%%%%%%%%%%%%%%%%%%%%%%%%%%%%%%%%%%%%%%%%%%%%%%%%%%%%%%%%%%%%%%%%%%%%%%%%%%%%%%%%%%%%%%%%%%%%%%%%%%%%%%%%%%
%\subsection{Pseudoreal representation and  $OSp(1|2)$ singlet}
%%%%%%%%%%%%%%%%%%%%%%%%%%%%%%%%%%%%%%%%%%%%%%%%%%%%%%%%%%%%%%%%%%%%%%%%%%
%%%%%%%%%%%%%%%%%%%%%%%%%%%%%%%%%%%%%%%%%%%%%%%%%%%%%%%%%%%%%%%%%%%%%%%%%%
%\subsubsection{$OSp(1|2)$ singlet}
%%%%%%%%%%%%%%%%%%%%%%%%%%%%%%%%%%%%%%%%%%%%%%%%%%%%%%%%%%%%%%%%%%%%%%%%%%%%%%%%%%%%%%%%%%%%%%%%%%%%%%%%%%%%%%%%%%%%%%%%%%%%%%%%%%%%%%%%%%%%%%%%%%%%
%%%%%%%%%%%%%%%%%%%%%%%%%%%%%%%%%%%%%%%%%%%%%%%%%%%%%%%%%%%%%%%%%%%%%%%%%
%%%%%%%%%%%%%%%%%%%%%%%%%%%%%%%%%%%%%%%%%%%%%%%%%%%%%%%%%%%%%%%%%%%%%%%%%%
%%%%%%%%%%%%%%%%%%%%%%%%%%%%%%%%%%%%%%%%%%%%%%%%%%%%%%%%%%%%%%%%%%%%%%%%%%
%\subsubsection{Pseudoreal representation}
%%%%%%%%%%%%%%%%%%%%%%%%%%%%%%%%%%%%%%%%%%%%%%%%%%%%%%%%%%%%%%%%%%%%%%%%%
The complex representation in $OSp(1|2)$ is related to the original by
 the unitary transformation,
$\tilde{l}_a=\mathcal{R}^{t}l_a \mathcal{R},~\tilde{l}_{\alpha}=\mathcal{R}^t l_{\alpha} \mathcal{R},$
where $\mathcal{R}$ is given by
%%%%%%%%%%%%%%%%%%%%%%%%%%%%%%%%%%%%%%%%%%%%%%%%%%%%%%%%%%%%%%%%%%%%%%%%%
\begin{equation}
\mathcal{R}=
\begin{pmatrix}
& 0 & 1 & 0 \\
& -1 & 0 & 0 \\
& 0 & 0 & 1 
\end{pmatrix}.
\label{mathcalR}\nonumber
\end{equation}
%%%%%%%%%%%%%%%%%%%%%%%%%%%%%%%%%%%%%%%%%%%%%%%%%%%%%%%%%%%%%%%%%%%%%%%%%%
Thus, the representation of $OSp(1|2)$ is  pseudo-real.
The properties of $\mathcal{R}$ are  as follows:  
$\mathcal{R}^t =\mathcal{R}^{\dagger}=\mathcal{R}^{\ddagger}=\mathcal{R}^{-1}$ 
 and $\mathcal{R}^2=(\mathcal{R}^t)^2=diag(-1,-1,1)$.
%%%%%%%%%%%%%%%%%%%%%%%%%%%%%%%%%%%%%%%%%%%%%%%%%%%%%%%%%%%%%%%%%%%%%%%%%%
Using the matrix $\mathcal{R}$, the complex spinor is given by  
%%%%%%%%%%%%%%%%%%%%%%%%%%%%%%%%%%%%%%%%%%%%%%%%%%%%%%%%%%%%%%%%%%%%%%%%%
$\tilde{\psi}=\mathcal{R}(\psi^{\ddagger})^{t}=
( v^{*}, -u^{*},\eta^{*})^t$.
%%%%%%%%%%%%%%%%%%%%%%%%%%%%%%%%%%%%%%%%%%%%%%%%%%%%%%%%%%%%%%%%%%%%%%%%%
Then,  without including the complex spinor $\psi^{\ddagger}$, the $OSp(1|2)$ singlet can be constructed from  $\psi$ and $\psi'$ alone:
%%%%%%%%%%%%%%%%%%%%%%%%%%%%%%%%%%%%%%%%%%%%%%%%%%%%%%%%%%%%%%%%%%%%%%%%%
${\tilde{\psi}}^{\ddagger}{\psi}'=\psi^t \mathcal{R}^{\ddagger}\psi'=(\psi')^t\mathcal{R}\psi.$
%%%%%%%%%%%%%%%%%%%%%%%%%%%%%%%%%%%%%%%%%%%%%%%%%%%%%%%%%%%%%%%%%%%%%%%%%%
%%%%%%%%%%%%%%%%%%%%%%%%%%%%%%%%%%%%%%%%%%%%%%%%%%%%%%%%%%%%%%%%%%%%%%%%%%%%%%%%%%%%%%%%%%%%%%%%%%%%%%%%%%%%%%%%%%%%%%%%%%%%%%%%%%%%%%%%%%%%%%%%%%%%
%\section{two particle state}
%%%%%%%%%%%%%%%%%%%%%%%%%%%%%%%%%%%%%%%%%%%%%%%%%%%%%%%%%%%%%%%%%%%%%%%%%%%%%%%%%%%%%%%%%%%%%%%%%%%%%%%%%%%%%%%%%%%%%%%%%%%%%%%%%%%%%%%%%%%%%%%%%%%%

We next study the two-particle state.
The total angular momenta of $OSp(1|2)$ are given by 
$L_a^{tot}=L_a(1)+L_a(2)$ and $L_{\alpha}^{tot}=L_{\alpha}(1)+L_{\alpha}(2),$
%%%%%%%%%%%%%%%%%%%%%%%%%%%%%%%%%%%%%%%%%%%%%%%%%%%%%%%%%%%%%%%%%%%%%%%%%%
where $L_a(i)$ and $L_{\alpha}(i)$ are 
the $OSp(1|2)$ generators of the $i$-th particle.
%%%%%%%%%%%%%%%%%%%%%%%%%%%%%%%%%%%%%%%%%%%%%%%%%%%%%%%%%%%%%%%%%%%%%%%%%%%%%%%%%%%%%%%%%%%%%%%%%%%%%%%%%%%%%%%%%%%%%%%%%%%%%%%%%%%%%%%%%%%%%%%%%%%%
%\subsection{Super coherent state}
%%%%%%%%%%%%%%%%%%%%%%%%%%%%%%%%%%%%%%%%%%%%%%%%%%%%%%%%%%%%%%%%%%%%%%%%%%%%%%%%%%%%%%%%%%%%%%%%%%%%%%%%%%%%%%%%%%%%%%%%%%%%%%%%%%%%%%%%%%%%%%%%%%%%%%%%%%%%%%%%%%%%%%%%%%%%%%%%%%%%%%%%%%%%%%%%%%%%%%%%%%%%%%%%%%%%%%%%%%%%%%
The two-particle supercoherent state 
located at the point $(\Omega_a,\Omega_{\alpha})$
  satisfies the  equation 
$[\Omega_a(\chi)L^{tot}_a+C_{\alpha\beta}\Omega_{\alpha}(\chi)L^{tot}_{\beta}] \psi_{\chi}^{(I,J)}=-J  \psi_{\chi}^{(I,J)}.$
%%%%%%%%%%%%%%%%%%%%%%%%%%%%%%%%%%%%%%%%%%%%%%%%%%%%%%%%%%%%%%%%%%%%%%%%%
The solution  is  written as
%%%%%%%%%%%%%%%%%%%%%%%%%%%%%%%%%%%%%%%%%%%%%%%%%%%%%%%%%%%%%%%%%%%%%%%%%%
\begin{equation}
\psi_{\chi}^{(I,J)}\!=\!(u_1v_2-v_1u_2-\eta_1\eta_2)^{I-J}\cdot\psi^{(J)}_{\chi}(u_1,\! v_1,\! \eta_1)\psi^{(J)}_{\chi}( u_2,\! v_2,\! \eta_2)\!,\nonumber
\end{equation}
%%%%%%%%%%%%%%%%%%%%%%%%%%%%%%%%%%%%%%%%%%%%%%%%%%%%%%%%%%%%%%%%%%%%%%%%%%
where the first component on the right-hand side is an $OSp(1|2)$ singlet
  that 
 determines the distance
 between  two particles.
 Thus, the two-particle state $\psi^{(I-J)}_{\chi}$ represents  an extended object whose spin is $J$, 
 whose center-of-mass is located at $(\Omega_a(\chi),\Omega_{\alpha}(\chi))$  and whose size is proportional to $(I-J)$. 
%%%%%%%%%%%%%%%%%%%%%%%%%%%%%%%%%%%%%%%%%%%%%%%%%%%%%%%%%%%%%%%%%%%%%%%%%
%%%%%%%%%%%%%%%%%%%%%%%%%%%%%%%%%%%%%%%%%%%%%%%%%%%%%%%%%%%%%%%%%%%%%%%%%%%%%%%%%%%%%%%%%%%%%%%%%%%%%%%%%%%%%%%%%%%%%%%%%%%%%%%%%%%%%%%%%%%%%%%%%%%
%\subsection{Super pseudopotential}
%%%%%%%%%%%%%%%%%%%%%%%%%%%%%%%%%%%%%%%%%%%%%%%%%%%%%%%%%%%%%%%%%%%%%%%%%%%%%%%%%%%%%%%%%%%%%%%%%%%%%%%%%%%%%%%%%%%%%%%%%%%%%%%%%%%%%%%%%%%%%%%%%%%%

In the LLL, this system  reduces to a  fuzzy supersphere 
\cite{hasebekimura0409230}.
Hence, the two-body interaction 
  is reduced to that on a  
fuzzy supersphere, which is expressed as a  supersymmetric extension of  Haldane's  pseudo-potential \cite{HaldanePRL51},  
$\Pi_I\cdot V(\Omega_a(1)\Omega_a(2)+C_{\alpha\beta}\Omega_{\alpha}(1)\Omega_{\beta}(2))\cdot\Pi_I\nonumber\\
=\!\!\!\sum_{J=0,1/2,\cdots,I}V_J^{(I)}P_J(L_a(1) L_a(2)+C_{\alpha\beta}L_{\alpha}(1)L_{\beta}(2)),$
where $P_J$ is the projection operator to the Hilbert space spanned by the states that form an irreducible representation of  $OSp(1|2)$ spin  $J$ .
It is noted that the $OSp(1|2)$ spin $J$ takes not only  integer values but also  half integer values. Physically, this implies  that  fermionic particles, as well as bosonic particles,   appear as  two-particle states 
 in  supersymmetric quantum Hall systems. 
%%%%%%%%%%%%%%%%%%%%%%%%%%%%%%%%%%%%%%%%%%%%%%%%%%%%%%%%%%%%%%%%%%%%%%%%%%

%%%%%%%%%%%%%%%%%%%%%%%%%%%%%%%%%%%%%%%%%%%%%%%%%%%%%%%%%%%%%%%%%%%%%%%%%%%%%%%%%%%%%%%%%%%%%%%%%%%%%%%%%%%%%%%%%%%%%%%%%%%%%%%%%%%%%%%%%%%%%%%%%%%
%\section{N-particle state}
%%%%%%%%%%%%%%%%%%%%%%%%%%%%%%%%%%%%%%%%%%%%%%%%%%%%%%%%%%%%%%%%%%%%%%%%%%%%%%%%%%%%%%%%%%%%%%%%%%%%%%%%%%%%%%%%%%%%%%%%%%%%%%%%%%%%%%%%%%%%%%%%%%%%
In Haldane's   work \cite{HaldanePRL51}, the 
Laughlin-Haldane wavefunction is constructed
 from the $SU(2)$ singlet state,
 which is apparently invariant under any rotations of the two-sphere.
Therefore, it is natural to  
use an $OSp(1|2)$ singlet wavefunction as 
  the supersymmetric extension of the  Laughlin-Haldane wavefunction.
 Explicitly, it is given by 
%%%%%%%%%%%%%%%%%%%%%%%%%%%%%%%%%%%%%%%%%%%%%%%%%%%%%%%%%%%%%%%%%%%%%%%%%
\begin{equation}
\Psi^{(m)}=\prod_{i<j}^{N} [-\psi_j^{t}\mathcal{R}\psi_i]^m
=\prod_{i<j}^N (u_iv_j-v_iu_j-\eta_i\eta_j)^m.\nonumber
\label{superLaughlin}
\end{equation}
%%%%%%%%%%%%%%%%%%%%%%%%%%%%%%%%%%%%%%%%%%%%%%%%%%%%%%%%%%%%%%%%%%%%%%%
When the product is expanded, it can be  easily seen that the sum of the powers of $u_i$ and $v_i$  is $m(N-1)$.
Also, the LLL constraint $m_1+m_2=N_{\Phi}$ for the monopole harmonics $u_{m_1,m_2}$  should  be satisfied.  
Hence, the  number of  particles and the number of magnetic cells are related as 
%%%%%%%%%%%%%%%%%%%%%%%%%%%%%%%%%%%%%%%%%%%%%%%%%%%%%%%%%%%%%%%%%%%%%%%%
$m(N-1)=N_{\Phi}.$
%%%%%%%%%%%%%%%%%%%%%%%%%%%%%%%%%%%%%%%%%%%%%%%%%%%%%%%%%%%%%%%%%%%%%%%%%
Thus, $\Psi^{(m)}$ describes a supersymmetric quantum Hall liquid at $\nu=1/m$ in the thermodynamic limit.
%The $OSp(1|2)$ transformations act on the supersphere as  ``rotations''.
%The many particles described by $\Psi^{(m)}$ are uniformly distributed on %the supersphere, because $\Psi^{(m)}$ is invariant under any 
%``rotation'' generated by $OSp(1|2)$. 
The supercoherent state for $\Psi^{(m)}$, whose center is 
 located at $(\Omega_a(\chi),\Omega_{\alpha}(\chi))$, is given by 
$\Psi^{(m)}_{\chi}=\prod_{i}^{N}(a^{*}u_i+b^{*}v_i-\xi^{*}\eta_i)^I\cdot 
\Psi^{(m)}.$
In fact, this wavefunction satisfies the supercoherent  equation,
$[\Omega_a(\chi)L_a+C_{\alpha\beta}\Omega_{\alpha}(\chi)L_{\beta}]\Psi^{(m)}_{\chi}=-\frac{I}{2}N \Psi^{(m)}_{\chi}.$

%%%%%%%%%%%%%%%%%%%%%%%%%%%%%%%%%%%%%%%%%%%%%%%%%%%%%%%%%%%%%%%%%%%%%%%%%%%%%%%%%%%%%%%%%%%%%%%%%%%%%%%%%%%%%%%%%%%%%%%%%%%%%%%%%%%%%%%%%%%%%%%%%%%%
%\subsection{Relation between  the Laughlin-Haldane wavefunction and  its 
%supersymmetric extension}
%%%%%%%%%%%%%%%%%%%%%%%%%%%%%%%%%%%%%%%%%%%%%%%%%%%%%%%%%%%%%%%%%%%%%%%%%%%%%%%%%%%%%%%%%%%%%%%%%%%%%%%%%%%%%%%%%%%%%%%%%%%%%%%%%%%%%%%%%%%%%%%%%%%%
The supersphere contains 
 an ordinary two-sphere as its ``body''.
The coordinates on this body-sphere are denoted  $\{y_a\}$.
 These satisfy $y_a^2=1$  and  are related to $\{x_a\}$ as
 $y_a=(1+\frac{1}{2}
\theta C\theta)x_a$.
Physical interpretations are easily formulated for phenomena on the body-sphere 
and some calculations are reduced to the ones on the body-sphere \cite{hasebekimura0409230}
. 
For instance, the supermagnetic flux integral   on 
the supersphere is reduced to the ordinary magnetic flux integral for 
  the Dirac monopole  
on the body-sphere. 
Thus, the body-sphere ``inside'' the supersphere has 
a Dirac  monopole at its center and 
  becomes a fuzzy sphere in the LLL.
Then, it is conjectured that   Haldane's original quantum Hall systems are realized on such a fuzzy body-sphere. 
The super Hopf spinor is rewritten 
$(u,~v,~\eta)=
(1-\frac{1}{4}\theta C\theta)
(\mu,~\nu,~\mu\theta_1+\nu\theta_2)$,
where 
$(\mu,~\nu)
= (\sqrt{\frac{1+y_3}{2}},~\frac{y_1+iy_2}{\sqrt{2(1+y_3)}})$
is a  Hopf spinor on the body-sphere \cite{hasebekimura0409230}. 
By inserting this form  into $\Psi^{(m)}$, 
  the supersymmetric extension of the Laughlin-Haldane wavefunction can be 
   expressed in terms of 
the body coordinates $\{y_a\}$ and soul coordinates $\{\theta_{\alpha}\}$ as
%%%%%%%%%%%%%%%%%%%%%%%%%%%%%%%%%%%%%%%%%%%%%%%%%%%%%%%%%%%%%%%%%%%%%%%%
\begin{align}
&\Psi^{(m)}\!\!=\!\!\prod_{i}^N \biggl(\! 1-\frac{m(N-1)}{4}\theta C\theta \! \biggr)_i\nonumber\\
&~~~~~~\cdot \prod_{i<j}\biggl(\! 1-m\frac{(\mu\theta_1+\nu\theta_2)_i(\mu\theta_1+\nu\theta_2)_j}{\mu_i\nu_j-\nu_i\mu_j} \!\biggr)\nonumber\\
&~~~~~~\cdot{\Phi}^{(m)},
\nonumber
\end{align}
%%%%%%%%%%%%%%%%%%%%%%%%%%%%%%%%%%%%%%%%%%%%%%%%%%%%%%%%%%%%%%%%%%%%%%%%% 
where ${\Phi}^{(m)}$ is the ordinary Laughlin-Haldane wavefunction on the body-sphere: 
%%%%%%%%%%%%%%%%%%%%%%%%%%%%%%%%%%%%%%%%%%%%%%%%%%%%%%%%%%%%%%%%%%%%%%%%%%
${\Phi}^{(m)}=\prod_{i<j}({\mu}_i{\nu}_j-{\nu}_i{\mu}_j)^m.$ 
%%%%%%%%%%%%%%%%%%%%%%%%%%%%%%%%%%%%%%%%%%%%%%%%%%%%%%%%%%%%%%%%%%%%%%%%%
The second term on the right-hand side 
%of Eq.(\ref{Laughlinrelation})
 yields nontrivial connections between the 
body-sphere  and the soul space.
Apparently, $\Psi^{(m)}$ is a singlet with respect to  $SU(2)$ transformations  
 generated by $\{L_a\}$, because it is a subgroup of $OSp(1|2)$.
However,
it is noted that, due to the second term, 
%in  Eq.(\ref{Laughlinrelation}),
 $\Psi^{(m)}$ does not possess the $SU(2)$ rotational symmetry of 
the body-sphere, which is generated by $\{(\mu~\nu)(-\frac{1}{2}\sigma_a^*)(\frac{\partial}{\partial\mu}~\frac{\partial}{\partial\nu})^t\}$.

%%%%%%%%%%%%%%%%%%%%%%%%%%%%%%%%%%%%%%%%%%%%%%%%%%%%%%%%%%%%%%%%%%%%%%%%%%%%%%%%%%%%%%%%%%%%%%%%%%%%%%%%%%%%%%%%%%%%%%%%%%%%%%%%%%%%%%%%%%%%%%%%%%%%
%%%%%%%%%%%%%%%%%%%%%%%%%%%%%%%%%%%%%%%%%%%%%%%%%%%%%%%%%%%%%%%%%%%%%%%%%%%%%%%%%%%%%%%%%%%%%%%%%%%%%%%%%%%%%%%%%%%%%%%%%%%%%%%%%%%%%%%%%%%%%%%%%%%%
%%%%%%%%%%%%%%%%%%%%%%%%%%%%%%%%%%%%%%%%%%%%%%%%%%%%%%%%%%%%%%%%%%%%%%%%%%
%%%%%%%%%%%%%%%%%%%%%%%%%%%%%%%%%%%%%%%%%%%%%%%%%%%%%%%%%%%%%%%%%%%%%%%%%%
%\subsection{Exact Hamiltonian}
%%%%%%%%%%%%%%%%%%%%%%%%%%%%%%%%%%%%%%%%%%%%%%%%%%%%%%%%%%%%%%%%%%%%%%%%%%
%%%%%%%%%%%%%%%%%%%%%%%%%%%%%%%%%%%%%%%%%%%%%%%%%%%%%%%%%%%%%%%%%%%%%%%%%%
Because, in $\Psi^{(m)}$, two particles $(i,j)$ have  a 
power at least  $m$,
%( The power of the two particles represents the distance between them.)
%Note that
 no two particles can become so close that we have $I-m < J_{ij}$.
Therefore, just as in the Laughlin-Haldane case \cite{HaldanePRL51},
 $\Psi^{(m)}$ is the exact ground state of the hard-core interaction Hamiltonian, 
%%%%%%%%%%%%%%%%%%%%%%%%%%%%%%%%%%%%%%%%%%%%%%%%%%%%%%%%%%%%%%%%%%%%%%%%%
$\Pi_{I}H_{m}^{int} \Pi_{I} =\sum_{i<j}\sum_{I-m<J}\!\!V_{J}P_{J}(L_a(i)L_a(j)+C_{\alpha\beta}L_{\alpha}(i)L_{\beta}(j)),$
%%%%%%%%%%%%%%%%%%%%%%%%%%%%%%%%%%%%%%%%%%%%%%%%%%%%%%%%%%%%%%%%%%%%%%%%%
with  energy $0$, where  $V_{J}> 0$.
This Hamiltonian is a direct generalization of  
Haldane's Hamiltonian, with  the  $OSp(1|2)$ Casimir operator replacing the $SU(2)$  one.
%%%%%%%%%%%%%%%%%%%%%%%%%%%%%%%%%%%%%%%%%%%%%%%%%%%%%%%%%%%%%%%%%%%%%%%%%
%\subsection{Stereographic projection}
%%%%%%%%%%%%%%%%%%%%%%%%%%%%%%%%%%%%%%%%%%%%%%%%%%%%%%%%%%%%%%%%%%%%%%%%%
%%%%%%%%%%%%%%%%%%%%%%%%%%%%%%%%%%%%%%%%%%%%%%%%%%%%%%%%%%%%%%%%%%%%%%%%%   
%%%%%%%%%%%%%%%%%%%%%%%%%%%%%%%%%%%%%%%%%%%%%%%%%%%%%%%%%%%%%%%%%%%%%%%%%
%\subsection{Filling fraction}
%%%%%%%%%%%%%%%%%%%%%%%%%%%%%%%%%%%%%%%%%%%%%%%%%%%%%%%%%%%%%%%%%%%%%%%%%
%%%%%%%%%%%%%%%%%%%%%%%%%%%%%%%%%%%%%%%%%%%%%%%%%%%%%%%%%%%%%%%%%%%%%%%%%%
%%%%%%%%%%%%%%%%%%%%%%%%%%%%%%%%%%%%%%%%%%%%%%%%%%%%%%%%%%%%%%%%%%%%%%%%
%%%%%%%%%%%%%%%%%%%%%%%%%%%%%%%%%%%%%%%%%%%%%%%%%%%%%%%%%%%%%%%%%%%%%%%%
%\section{super Hall current}
%%%%%%%%%%%%%%%%%%%%%%%%%%%%%%%%%%%%%%%%%%%%%%%%%%%%%%%%%%%%%%%%%%%%%%%%%%%%%%%%%%%%%%%%%%%%%%%%%%%%%%%%%%%%%%%%%%%%%%%%%%%%%%%%%%%%%%%%%%%%%%%%%%%%
%%%%%%%%%%%%%%%%%%%%%%%%%%%%%%%%%%%%%%%%%%%%%%%%%%%%%%%%%%%%%%%%%%%%%%%%%%

Due to the noncommutative algebra on the fuzzy supersphere, the super Hall currents
 $I_a  = \frac{d}{dt}x_a $ and  $I_{\alpha}=\frac{d}{dt}\theta_{\alpha} $
 can be expressed as  
%%%%%%%%%%%%%%%%%%%%%%%%%%%%%%%%%%%%%%%%%%%%%%%%%%%%%%%%%%%%%%%%%%%%%%%%%
\begin{align}
&I_a=-i[x_a,V]=\alpha\epsilon_{abc}x_b E_c -i\alpha\frac{1}{2}(\sigma_a C)_{\alpha\beta}\theta_{\alpha}E_{\beta},   \nonumber\\
&I_{\alpha}=-i[\theta_{\alpha},V]=-i\alpha\frac{1}{2} x_a(\sigma_a)_{\beta\alpha}E_{\beta}-i\alpha\frac{1}{2} \theta_{\beta}(\sigma_a)_{\beta\alpha}E_a,\nonumber
\label{supercurrent}
\end{align}
%%%%%%%%%%%%%%%%%%%%%%%%%%%%%%%%%%%%%%%%%%%%%%%%%%%%%%%%%%%%%%%%%%%%%%%%%%
where  $\alpha=2/I$, the superelectric fields $E_a$ and $E_{\alpha}$ 
are defined as 
%%%%%%%%%%%%%%%%%%%%%%%%%%%%%%%%%%%%%%%%%%%%%%%%%%%%%%%%%%%%%%%%%%%%%%%%%% 
$E_a=-{\partial_a}V$ and $ E_{\alpha}=-C_{\alpha\beta}{\partial_{\beta}}V$,
%%%%%%%%%%%%%%%%%%%%%%%%%%%%%%%%%%%%%%%%%%%%%%%%%%%%%%%%%%%%%%%%%%%%%%%%%% 
and we have used the fact that  the  
Hamiltonian is reduced to the potential term $V$ in the LLL.
It  can be  easily seen that the super Hall currents satisfy the equation
%%%%%%%%%%%%%%%%%%%%%%%%%%%%%%%%%%%%%%%%%%%%%%%%%%%%%%%%%%%%%%%%%%%%%%%%%%
%\begin{equation}
$E_a I_a+C_{\alpha\beta}E_{\alpha}I_{\beta}=0,$
%\label{Hallrelation}
%\end{equation}
%%%%%%%%%%%%%%%%%%%%%%%%%%%%%%%%%%%%%%%%%%%%%%%%%%%%%%%%%%%%%%%%%%%%%%%%%
which expresses the orthogonality of the Hall currents and the electric fields in the supersymmetric sense.
%%%%%%%%%%%%%%%%%%%%%%%%%%%%%%%%%%%%%%%%%%%%%%%%%%%%%%%%%%%%%%%%%%%%%%%%%%%%%%%%%%%%%%%%%%%%%%%%%%%%%%%%%%%%%%%%%%%%%%%%%%%%%%%%%%%%%%%%%%%%%%%%%%%%
%\section{excited states}
%%%%%%%%%%%%%%%%%%%%%%%%%%%%%%%%%%%%%%%%%%%%%%%%%%%%%%%%%%%%%%%%%%%%%%%%%%%%%%%%%%%%%%%%%%%%%%%%%%%%%%%%%%%%%%%%%%%%%%%%%%%%%%%%%%%%%%%%%%%%%%%%%%%% 
When we pierce (eliminate) the supermagnetic flux at some point on the supersphere adiabatically, 
 superelectric fields are induced and circle the supermagnetic flux,
 as described by  the super Faraday law.
Due  to the Hall orthogonality discussed above,
 super Hall currents flow radially from the point at  which the 
supermagnetic flux is pierced (eliminated).
As a consequence, a charged deficit (excess), which we call a quasi-hole (quasi-particle), is generated at this point.  
The corresponding operators  at the point 
$(\Omega_a(\chi),\Omega_{\alpha}(\chi))$  are   given by 
%%%%%%%%%%%%%%%%%%%%%%%%%%%%%%%%%%%%%%%%%%%%%%%%%%%%%%%%%%%%%%%%%%%%%%%%%%
\begin{align}
&A^{\ddagger}(\chi)=\prod_{i}^N\psi^t_i \mathcal{R}^t \chi
%=\prod_{i}^N(b u_i-a v_i+\xi\eta_i) 
~~\text{(quasi-hole)},\nonumber\\
&A(\chi)=\prod_i^N \chi^{\ddagger}\mathcal{R}\frac{\partial}{\partial\psi_i}
%=\prod_{i}^N(b^{*} \frac{\partial}{\partial u_i}-
%a^{*}\frac{\partial}{\partial v_i}+\xi^{*}\frac{\partial}%{\partial \eta_i})
~~ \text{(quasi-particle)}.\nonumber
\end{align}
%%%%%%%%%%%%%%%%%%%%%%%%%%%%%%%%%%%%%%%%%%%%%%%%%%%%%%%%%%%%%%%%%%%%%%%%%%
The  quasi-hole operator satisfies the commutation relation
%%%%%%%%%%%%%%%%%%%%%%%%%%%%%%%%%%%%%%%%%%%%%%%%%%%%%%%%%%%%%%%%%%%%%%%%%%
%\begin{equation}
$[\Omega_a(\chi)L_a+C_{\alpha\beta}\Omega_{\alpha}(\chi)L_{\beta},A^{\ddagger}(\chi)]=\frac{N}{2}A^{\ddagger}(\chi).$
%\end{equation}
%%%%%%%%%%%%%%%%%%%%%%%%%%%%%%%%%%%%%%%%%%%%%%%%%%%%%%%%%%%%%%%%%%%%%%%%%%
Thus, when a  quasi-hole (quasi-particle) is created, 
the angular momentum in the direction of the point $(\Omega_a,\Omega_{\alpha})$
increases (decreases) by $\frac{1}{2}N$.
The statistics of the  quasi-particles are bosonic, i.e., 
%%%%%%%%%%%%%%%%%%%%%%%%%%%%%%%%%%%%%%%%%%%%%%%%%%%%%%%%%%%%%%%%%%%%%%%%%%
$[A^{\ddagger}(\chi),A^{\ddagger}(\chi')]=[A(\chi),A(\chi')]=0$ and 
$[A(\chi),A^{\ddagger}(\chi)]=1.$
%%%%%%%%%%%%%%%%%%%%%%%%%%%%%%%%%%%%%%%%%%%%%%%%%%%%%%%%%%%%%%%%%%%%%%%%%
%%%%%%%%%%%%%%%%%%%%%%%%%%%%%%%%%%%%%%%%%%%%%%%%%%%%%%%%%%%%%%%%%%%%%%%%%%%%%%%%%%%%%%%%%%%%%%%%%%%%%%%%%%%%%%%%%%%%%%%%%%%%%%%%%%%%%%%%%%%%%%%%%%%%
%\subsection{Fractionally charged  quasi-particles}
%%%%%%%%%%%%%%%%%%%%%%%%%%%%%%%%%%%%%%%%%%%%%%%%%%%%%%%%%%%%%%%%%%%%%%%%%%%%%%%%%%%%%%%%%%%%%%%%%%%%%%%%%%%%%%%%%%%%%%%%%%%%%%%%%%%%%%%%%%%%%%%%%%%%
For $\nu=1/m$, the number of  particles and  the number of  supermagnetic 
fluxes are related as   
%%%%%%%%%%%%%%%%%%%%%%%%%%%%%%%%%%%%%%%%%%%%%%%%%%%%%%%%%%%%%%%%%%%%%%%%%
$N_{\Phi}=m(N-1)$,
%%%%%%%%%%%%%%%%%%%%%%%%%%%%%%%%%%%%%%%%%%%%%%%%%%%%%%%%%%%%%%%%%%%%%%%%%
whose variation reads 
%%%%%%%%%%%%%%%%%%%%%%%%%%%%%%%%%%%%%%%%%%%%%%%%%%%%%%%%%%%%%%%%%%%%%%%%%
$\delta N=\frac{1}{m}\delta N_{\Phi}$.
%%%%%%%%%%%%%%%%%%%%%%%%%%%%%%%%%%%%%%%%%%%%%%%%%%%%%%%%%%%%%%%%%%%%%%%%%
Because each  quasi-particle corresponds to a  supermagnetic flux,
this relation implies that the charge of a quasi-particle is 
fractional, namely, $e^*=1/m$, as in the original case.
However, it is noted that in the present case the 
 charge deficit is made by  both the bosonic current $I_a$ and the fermionic current $I_{\alpha}$ on the supersphere.
%%%%%%%%%%%%%%%%%%%%%%%%%%%%%%%%%%%%%%%%%%%%%%%%%%%%%%%%%%%%%%%%%%%%%%%%%%%%%%%%%%%%%%%%%%%%%%%%%%%%%%%%%%%%%%%%%%%%%%%%%%%%%%%%%%%%%%%%%%%%%%%%%%%
%\section{Super Haldane-Halperin Hierarchy}
%%%%%%%%%%%%%%%%%%%%%%%%%%%%%%%%%%%%%%%%%%%%%%%%%%%%%%%%%%%%%%%%%%%%%%%%%%%%%%%%%%%%%%%%%%%%%%%%%%%%%%%%%%%%%%%%%%%%%%%%%%%%%%%%%%%%%%%%%%%%%%%%%%%

On the body-sphere, the supersymmetric quantum Hall liquid reduces to an ordinary quantum Hall liquid and  the Haldane-Halperin hierarchy is realized.
Because the behavior on the supersphere 
 is mapped to that on the body-sphere,
  it is natural to conjecture that the supersymmetric extension of the  
 Haldane-Halperin hierarchy is realized on the supersphere, where the 
super quasi-particles repeatedly condense to form a hierarchy of    supersymmetric quantum Hall liquids.

%%%%%%%%%%%%%%%%%%%%%%%%%%%%%%%%%%%%%%%%%%%%%%%%%%%%%%%%%%%%%%%%%%%%%%%%%%
%%%%%%%%%%%%%%%%%%%%%%%%%%%%%%%%%%%%%%%%%%%%%%%%%%%%%%%%%%%%%%%%%%%%%%%%%%
%\section{summary and discussion}
%%%%%%%%%%%%%%%%%%%%%%%%%%%%%%%%%%%%%%%%%%%%%%%%%%%%%%%%%%%%%%%%%%%%%%%%%%
%%%%%%%%%%%%%%%%%%%%%%%%%%%%%%%%%%%%%%%%%%%%%%%%%%%%%%%%%%%%%%%%%%%%%%%%%%
\vspace{0.3cm}
To summarize, we have constructed a low-dimensional supersymmetric quantum Hall system
 and investigated its basic properties. 
Similar to the original quantum Hall systems, this system raises many  interesting questions, for instance,  those regarding edge excitations, 
 anyonic objects,  
 effective field theory,
 and the relation of the present system to  integrable models.
In a planar limit, this system  reduces to a supersymmetric harmonic oscillator system \cite{Hasebeappear}, which is related to the Pauli Hamiltonian for a spin-$1/2$ particle with gyromagnetic factor $2$ or the Jaynes-Cummings model without interaction terms  used in quantum optics. 
There are many real systems which show  supersymmetric properties \cite{Junker}.
It would be worthwhile to investigate the realization of the   supersymmetric  holonomy, like the Wilczek-Zee nonabelian holonomy (which  generally appear in the presence of  degenerate energy levels) and  possible relevance to the supersymmetric quantum Hall effects, 
  in  real systems.  
%
%The supersymmetric quantum Hall effects would be relevant in such real %systems just like the spin-Hall effects, which  occur in real systems 
%due to the ``magnetic monopole'' in the momentum space.
With regard to  high energy physics, one of the most important task is 
 to  develop a higher-dimensional generalization   based on supersymmetric 2-nd and 3-rd Hopf maps.  
Once these are constructed, we should proceed to the investigation of their relation  to D-brane systems and super twistor models.
%It is expected that the supersymmetric version of the Myers effects can %be seen.
%Further, such systems may yield a model for super gravity as their edge %excitations. 

%%%%%%%%%%%%%%%%%%%%%%%%%%%%%%%%%%%%%%%%%%%%%%%%%%%%%%%%%%%%%%%%%%%%%%%%%%%\section*{ACKNOWLEDGEMENTS}
\vspace{0.3cm}
I would like to acknowledge 
Yusuke Kimura for useful discussions.
I am also glad to thank Taichiro Kugo, Hiroshi Kunitomo, Naoki Sasakura
 and Tatsuya Tokunaga 
 for conversations. 
This work was supported by a JSPS fellowship. 
%%%%%%%%%%%%%%%%%%%%%%%%%%%%%%%%%%%%%%%%%%%%%%%%%%%%%%%%%%%%%%%%%%%%%%%%%%

\end{document}